\documentclass[12pt]{article}

\usepackage{graphicx}
\usepackage{color}

\begin{document}
\baselineskip=25pt

\vspace{0.1cm}
\begin{center}
{\bf Physical realization and possible identification of topological excitations in quantum Heisenberg ferromagnet on  lattice}
\end{center}
\begin{center}
\vspace{0.5cm}
{ Ranjan Chaudhury}$^a$ and { Samir K. Paul}$^b$\\
$S.~ N.~ Bose~ National~ Centre~ For~ Basic~ Sciences , ~~ Block$-$JD,~ Sector$-$III , ~ Salt~ Lake$\\
$Calcutta$-$700098,~India$
\end{center}
\vspace{0.5cm}

a)    {$ranjan_{021258}@yahoo.com$ , ranjan@boson.bose.res.in}

b)    { smr@boson.bose.res.in}

\vspace{1.00cm}

Physical configurations corresponding to topological excitations present in the XY limit of a quantum spin $\frac {1}{2}$ Heisenberg ferromagnet, are investigated on a two dimensional square lattice .  Quantum vortices (anti-vortices) are constructed in terms of coherent spin field components and the crucial role of the associated Wess-Zumino term is highlighted . It is shown that this term can identify a large class of vortices (anti-vortices). In particular the  excitations with odd topological charge belonging to this class , are found to exhibit a self-similar pattern regarding the internal charge distribution . Our formalism is distinctly different from the conventional approach for the  construction of quantum vortices ( anti-vortices ).

\vspace{0.2cm}

\vspace{10.00cm}

{\bf Introduction}

\vspace{0.1cm}
The physical existence of topological excitations in quantum ferromagnetic spin systems in two dimension can be investigated by making use of  coherent state formulation $^{1,2,3}$. The XY-limit of  Heisenberg ferromagnet  also belongs to this class of systems .  It has been shown that in the case of a Heisenberg antiferromagnetic chain, the Wess-Zumino ( WZ) term in the effective action  is genuinely a topological term $^{1}$. Moreover in the long wavelength limit the explicit expression of this term is similar to that of the topological term for a nonlinear sigma model. Extending this approach further, it was shown by us $^{3}$  that it is possible to express the WZ term as a topological term, also in the case of a ferromagnetic chain in the long wavelength limit. We generalized this to the   case of a 2D ferromagnet $^{3,4}$ to demonstrate that in the above limit we could indeed get an expression from the WZ term , indicating the possibility of topological excitations.\\  
 
 Thus at this stage of analysis it is necessary to examine the physical configurations of these excitations in terms of coherent fields as well as the consequence of the WZ term, explicitly on a two dimensional square lattice. At the moment, we do this for the XY-limit of the 2D Heisenberg spin $\frac {1}{2}$ ferromagnet. We show that the WZ term really identifies a large class of the  topological excitations . Furthermore we demonstrate that this term can clearly  differentiate between vortices ( and antivortices ) with different charges within this class. It is mentionworthy that our entire analysis is valid for all temperatures . \\
{\bf Mathematical Formulation}\\
  The quantum Euclidean action ${S_E}[{\bf n}]$ for the spin coherent fields 
${\bf n}(t)$ can be written as $^{2,3}$ \\
\begin{equation}
S_E[{\bf n}] = -isS_{WZ}[{\bf n}]+{\frac{s{\delta t}}{4}}{\int_0^{\beta}}dt{{[{\partial_t}{{\bf n}(t)}]}^2} + {\int_0^{\beta}}dt H({\bf n})     
\end{equation}       
where $s$ is the magnitude of the spin ($s = {\frac{1}{2}}$ in the present case ) and 
\begin{equation}
H({\bf n})=\langle {\bf n}\vert H({\bf S})\vert {\bf n}\rangle
\end{equation}
$H({\bf S})$ being the spin Hamiltonian in representation '$s$'.
In the limit $\delta t\rightarrow 0$ the term ${\frac{s{\delta t}}{4}}{\int_0^{\beta}}dt{{[{\partial_t}{{\bf n}(t)}]}^2}$ on the right hand side of $Eqn.(1)$ vanishes and henceforth we omit this term in the expression for the action.The Wess-Zumino term $S_{WZ}$ is given by$^{2}$
\begin{equation}
{S_{WZ}}[{\bf n}]={\int_0^{\beta}}dt{\int_0^1}d{\tau} {\bf n}(t,{\tau})\cdot {\partial_t}{\bf n}(t,{\tau})\wedge {\partial_{\tau}}{\bf n}(t,{\tau})
\end{equation}
with ${\bf n}(t,0)\equiv {\bf n}(t)$, ${\bf n}(t,1)\equiv {{\bf n}_0}$, and
${\bf n}(0,{\tau})\equiv {\bf n}(\beta ,\tau )$, $t\in [0,\beta ]$, $\tau \in [0,1]$ , $\beta$ being the usual inverse temperature.\\
The expression in equation(3) is the area of the cap bounded by the trajectory  $\Gamma$ parametrized by ${\bf n}(t)$ $[{\equiv} ({n_1}(t),{n_2}(t),{n_3}(t))]$  on the sphere:   
\begin{equation}
{\bf n}\cdot {\bf n} = 1
\end{equation}

Here $\vert \bf n \rangle$ is the spin coherent state on a single lattice point as defined in Refs. 1-3 and all the expressions above are on a single lattice point. We calculate the difference (${\delta}{S_{WZ}}$) between the WZ terms on a pair of neighbouring lattice points$^2$  and express the WZ term for the whole lattice as 
\begin{eqnarray}
{S_{WZ}^{tot}} & = &{\sum_{i,j=1}^{2N}} {S_{WZ}}[{\bf n}(ia,ja)]\nonumber\\
& = &{\sum_{i,j=1}^{2N}} [{\frac{1}{2}}{S_{WZ}}[{\bf n}((i-1)a,ja)] + {\frac{1}{2}}{S_{WZ}}[{\bf n}(ia,(j-1)a)]\nonumber\\ 
&   & +{\frac{1}{2}}{\delta_x}{S_{WZ}}[{\bf n}(ia,ja)] +{\frac{1}{2}}{\delta_y}{S_{WZ}}[{\bf n}(ia,ja)]] 
\end{eqnarray}
where
\begin{eqnarray}  
{\delta_x}{S_{WZ}}[{\bf n}(\bf r)] & = & {S_{WZ}}[{\bf n}(ia,ja)] - {S_{WZ}}[{\bf n}((i-1)a,ja)]\nonumber\\ 
& = &{\int_0^{\beta}}dt [{\delta_x} {\bf n}\cdot ({\bf n}\wedge {\partial_t}{\bf n})]({\bf r})\nonumber\\ 
{\delta_y}{S_{WZ}}[{\bf n}(\bf r)] & = & {S_{WZ}}[{\bf n}(ia,ja)] - {S_{WZ}}[{\bf n}(ia,(j-1)a)]\nonumber\\ 
 & = &{\int_0^{\beta}}dt[{\delta_y} {\bf n}\cdot ({\bf n}\wedge {\partial_t}{\bf n})]({\bf r})    
\end{eqnarray}
 ${\bf r}=(ia,ja)$ , a being the lattice spacing and $i,j = 1,2,.....,2N$.
Now the spin Hamiltonian corresponding to anisotropic Heisenberg ferromagnet of $XXZ$ type is given by
\begin{equation}
H({\bf S}) = - g{\sum_{\langle {\bf r},{\bf r}\rangle}}{\bf {\tilde S}}({\bf r})\cdot {\bf {\tilde S}}({\bf {r\prime}}) - g{\lambda_z}{\sum_{\langle {\bf r},{\bf {r\prime}}\rangle}}{S_z}({\bf r}){S_z}({\bf {r\prime}})
\end{equation} 
with $g\ge 0$ and $0\le {\lambda_z}\le 1$ , $\bf r$,$\bf {r\prime}$ run over the lattice, and $\langle {\bf r},{\bf {r\prime}}\rangle$ signifies nearest neighbours  and ${\bf S} = ({\bf {\tilde S}}, S_z)$.\\
It follows from Equations $(2)$ and $(7)$ that the spin Hamiltonian in terms of coherent fields is given by 
\begin{equation}
H({\bf n}) = -g{\sum_{\langle (i,j),(r,s)\rangle}}{\bf {\tilde n}}(ia,ja)\cdot {\bf {\tilde n}}(ra,sa) - {\lambda_z}{\sum_{\langle (i,j),(r,s)\rangle}}{n_z}(ia,ja) {n_z}(ra,sa)   
\end{equation}
Now the quantum action (Euclidean)for the two dimensional anisotropic ferromagnet is given by
\begin{eqnarray}
{S_E}[{\bf n}]& = &-is{\sum_{i,j}}{S_{WZ}}[{\bf n}(ia,ja)] + {\int_0^{\beta}}dt [-g{s^2}{\sum_{\langle (i,j),(r,s)\rangle}}{\bf {\tilde n}}(i,j)\cdot{\bf {\tilde n}}(r,s) \nonumber\\
&  & - g{\lambda_z}{s^2}{\sum_{\langle (i,j),(r,s)\rangle}}{n_z}(ia,ja){n_z}(ra,sa)]
\end{eqnarray}
where $s=\frac {1}{2}$ since we are considering extreme quantum case.\\
In order to evaluate WZ-term [ i.e.,${S_{WZ}^{tot}}$]  on a vortex plaquette, we need to know the expression of the WZ-term on the lattice with the help of $Eqs.(5)$  \& $(6)$.\\ 
Now
\begin{eqnarray}
{\delta_x}{\bf n}(ia,ja) & = &{\bf n}( ia,ja ) - {\bf n}( (i-1)a,ja )\nonumber\\
{\delta_y}{\bf n}(ia,ja) & = &{\bf n}( ia,ja ) - {\bf n}( ia,(j-1)a )
\end{eqnarray} 
From Eqns. (5) , (6) and (10) we get
\begin{eqnarray}
2{S_{WZ}}[{\bf n}(ia,ja)]& = &{S_{WZ}}[{\bf n}( (i-1)a, ja )] + {S_{WZ}}[{\bf n}( ia, (j-1)a )]\nonumber\\
&  &+{\int_0^{\beta}}dt [{\bf n}(ia,ja)-{\bf n}((i-1)a,ja)]\cdot ({\bf n}\wedge {\partial_t}{\bf n})(ia,ja)\nonumber\\
&  &+{\int_0^{\beta}}dt [{\bf n}(ia,ja)-{\bf n}(ia, (j-1)a)]\cdot ({\bf n}\wedge {\partial_t}{\bf n})(ia,ja)\nonumber\\
& = &{S_{WZ}}[{\bf n}( (i-1)a, ja )] + {S_{WZ}}[{\bf n}( ia, (j-1)a )]          \nonumber\\
&   & -{\int_0^{\beta}}dt {\bf n}( (i-1)a,ja )\cdot ({\bf n}\wedge {\partial_t}{\bf n})( ia,ja )\nonumber\\
&   &-{\int_0^{\beta}}dt  {\bf n}( ia,(j-1)a )\cdot ({\bf n}\wedge {\partial_t}{\bf n})(ia,ja)
\end{eqnarray}
The time derivative of the coherent spin field, $\partial_t{\bf n}$ in the above equation can be expressed in terms of coherent fields and their spatial derivatives through equations of motion. This we obtain from the action given by $Eq.(9)$ in the continuum limit as an approximation . Thus in this limit we have,
\begin{equation}
{S_{tot}} = {S_E}[{\bf n}] + {\int} {d^2}x {\int_0^{\beta}} {\lambda} ({\bf x}, t)({{\bf n}^2}({\bf x},t) - 1)
\end{equation}
where the first action in the right hand side of the above equation is the continuum version of the action given by $Eq.(9)$ . ${\lambda}({\bf x},t)$ in the above equation is an auxiliary field playing the role of a multiplier . The equations which follow from the above action are the following
\begin{eqnarray}
{\frac{i}{2g{a^2}}}{\partial_t}{n_1}&  = & {({\bf n}\wedge {{\bf{\nabla}}^2}{\bf n})_1} - (1- {\lambda_z}){n_2}{{\bf{\nabla}}^2}{n_3}\nonumber\\
{\frac{i}{2g{a^2}}}{\partial_t}{n_2}&  = & {({\bf n}\wedge {{\bf{\nabla}}^2}{\bf n})_2} + (1- {\lambda_z}){n_1}{{\bf{\nabla}}^2}{n_3}\nonumber\\
{\frac{i}{2g{a^2}}}{\partial_t}{n_3}&  = & {({\bf n}\wedge {{\bf{\nabla}}^2}{\bf n})_3} 
\end{eqnarray}   
Now we substitute the expressions for ${\partial_t}{\bf n}$ from the above equations, into $Eq. (11)$ with discretized version of the derivatives on the right hand side of $Eq.(13)$ . This enables us to obtain an expression for  ${S_{WZ}^{tot}}$  corresponding to the $XXZ$ Heisenberg ferromagnet.  The  expressions of the derivatives to be used on the two dimensional lattice are as follows [assuming the common factor ${\frac{i}{2g{a^2}}}$ to be equal to 1 for simplicity]   :-
\begin{eqnarray}
{\partial_x}{\bf n}(ia,ja) & = &{\frac{1}{a}}[{\bf n}(ia,ja) - {\bf n}((i-1)a,ja)]\nonumber\\
{\partial_y}{\bf n}(ia,ja) & = &{\frac{1}{a}}[{\bf n}(ia,ja) - {\bf n}(ia,(j-1)a)]
\end{eqnarray}
\begin{eqnarray}
{{\bf \nabla}^2}{\bf n}(ia,ja)& = & {\frac{1}{a^2}}[2{\bf n}(ia,ja) - 2{\bf n}((i-1)a,ja)- 2{\bf n}(ia,(j-1)a)\nonumber\\
&   & + {\bf n}((i-2)a,ja) + {\bf n}(ia,(j-2)a)]
\end{eqnarray}
To keep the calculations simple but reasonable , we retain the intra-plaquette contributions by imposing a "local periodic boundary condition" (local PBC) viz.,
\begin{eqnarray}
{\bf n}(ia , ja)& = & {\bf n}(ia+(Q+1)ia , ja) \nonumber\\
{\bf n}(ia , ja)& = & {\bf n}(ia , (Q+1)ja+ja)
\end{eqnarray}
where $Q$ is the "topological charge" of the plaquette under consideration. We use the above condition in the following section where we consider vortices with topological charges 1,2,3 etc. Henceforth we denote a vortex with topological charge $Q$ ( a configuration on the lattice in which the spin undergoes a $2\pi \cdot Q$ rotation as we go around the plaquette once ) as $Q$-$vortex$ .

\vspace{0.5cm}

{\bf Calculations and Results}\\
:{\bf Analysis of 1-vortex} :\\
\begin{figure}[!htbp]
\begin{center}
\includegraphics[keepaspectratio,width=7cm]{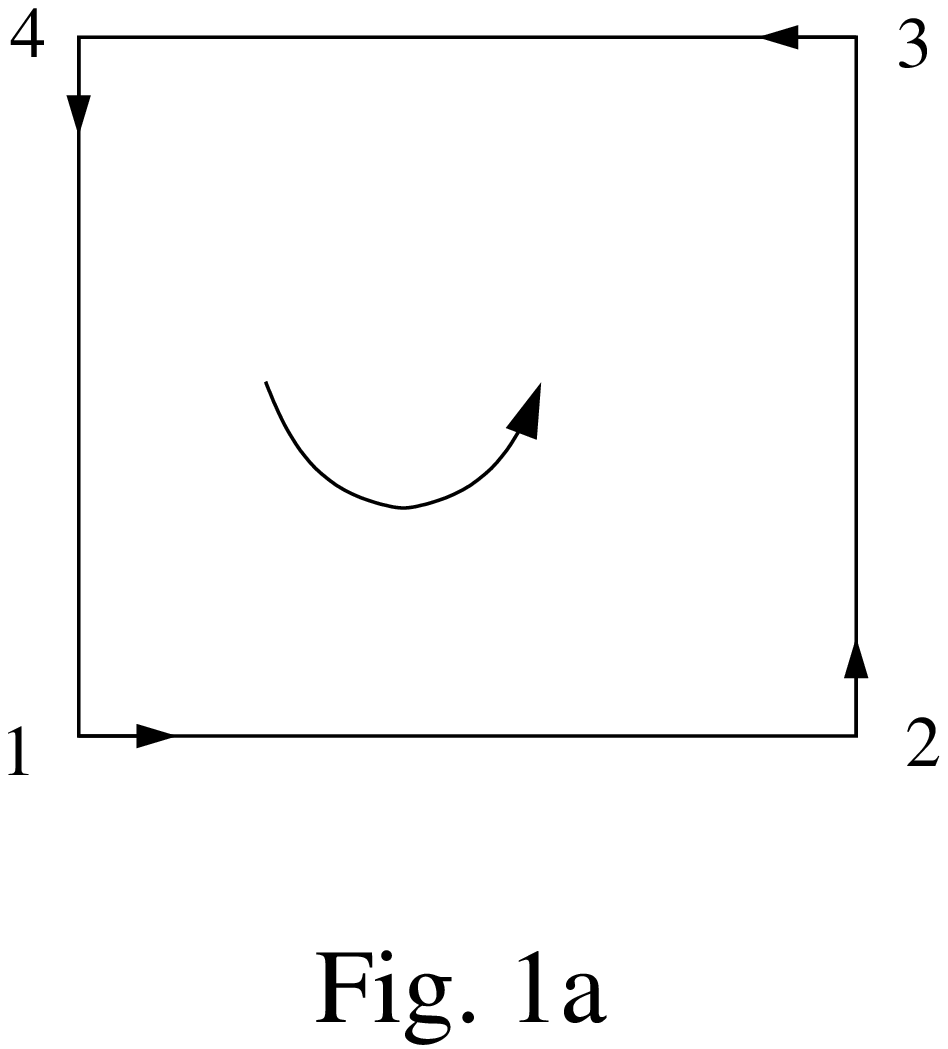}
\includegraphics[keepaspectratio,width=7cm]{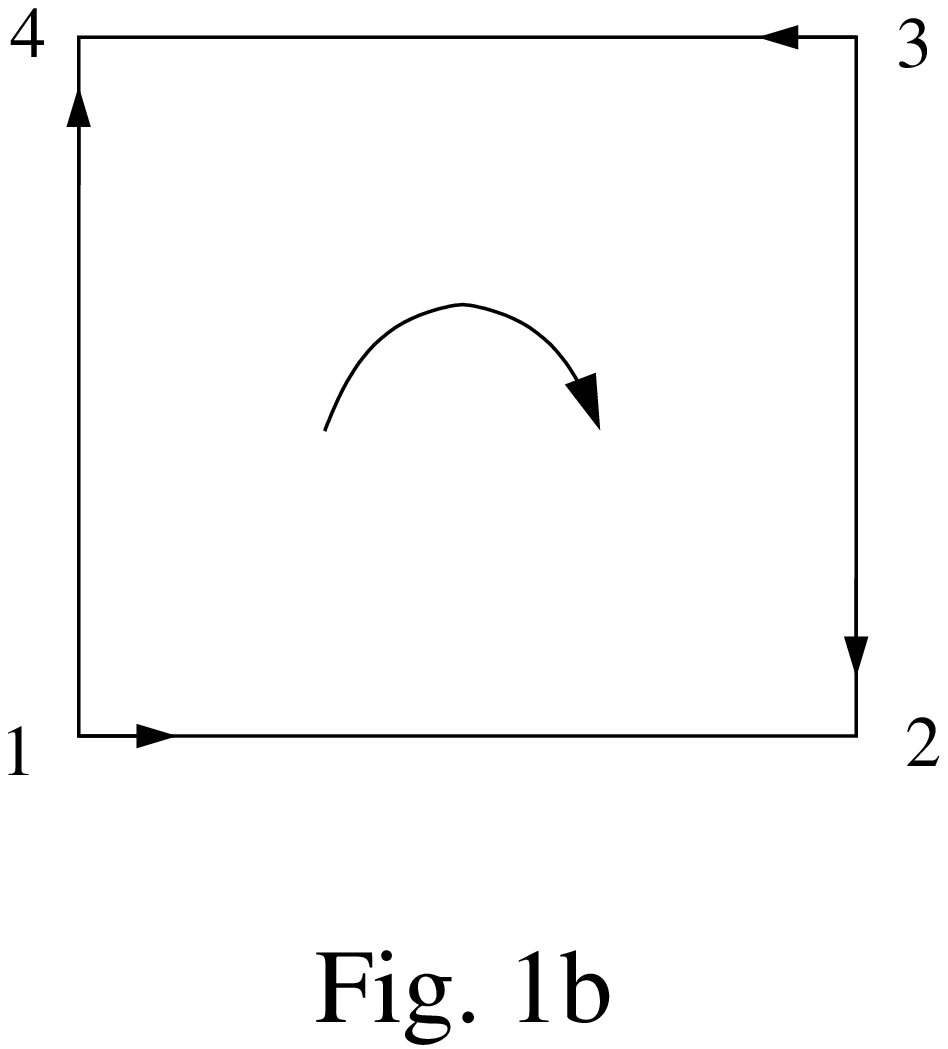}
\end{center}
\caption{(a) 1-vortex , (b) 1-anti-vortex}
\end{figure}
Eq.(11) shows that the last two terms are the contributions  from ${S_{WZ}}(ia,ja)$ to the vertices   2 , 3  and 4 with coordinates  $(ia,(j-1)a)$ , $(ia,ja)$  and $((i-1)a,ja)$ respectively , belonging to the 1-vortex [$see~Fig.1$]. Other contributions to the plaquette can be obtained by considering  ${S_{WZ}}$ at different lattice points.
 In other words , for a vortex ( anti-vortex ) plaquette having vertices 1 , 2 , 3 and 4 with coordinates $((i-1)a,(j-1)a)$, $(ia,(j-1)a)$, $(ia,ja)$ and $((i-1)a,ja)$ respectively,  $S_{WZ}$  associates the  kind of terms, similar to the ones discussed above , to a 1-vortex ( 1-antivortex ) [see${Fig.1}$]. It is easy to check that the contribution from  $S_{WZ}$  to a single plaquette (with vertices  $((i-1)a,(j-1)a)$, $(ia,(j-1)a)$,  $(ia,ja)$  and $((i-1)a,ja)$ are given by (see $Fig.1$ )
\begin{eqnarray}
{V_1} & = & C \{ {\int_0^{\beta}}dt [{\bf n}((i-1)a,ja)\cdot ({\bf n}\wedge {\partial_t}{\bf n}) (ia,ja)\nonumber\\
&  & +{\bf n}(ia,(j-1)a)\cdot ({\bf n}\wedge {\partial_t}{\bf n}) (ia,ja)\nonumber\\
&  & + {\bf n}((i-1)a,(j-1)a)\cdot ({\bf n}\wedge {\partial_t}{\bf n}) (ia,(j-1)a)\nonumber\\
&  &+{\bf n}((i-1)a,(j-1)a)\cdot ({\bf n}\wedge {\partial_t}{\bf n}) ((i-1)a,ja)
] \}
\end{eqnarray}
where the proportionality constant $C$ depends on the size of the lattice . Henceforth we drop $C$ from all the expressions for convenience. Since we will be interested in the extreme $XY$- anisotropic limit we we assume ${n_3} = \sin \epsilon$ on each lattice point where $\epsilon$ is very small. In this equation one of the components ${n_1}$ , ${n_2}$  is given by $1-\delta$ so that the other is given by $\pm {\sqrt {{\cos}^2}\epsilon - {{(1-\delta )}^2}}$ , $\delta$ being small . In Figure-1 we have a quantum vortex in which the horizontal arrow $\rightarrow$ at a vertex implies  $n_1$ with value $1-\delta$ and the vertical arrow $\uparrow$ implies $n_2$ with value $1-\delta$. Also in this figure the horizontal arrow $\leftarrow$ at a vertex implies $n_1$ with value $-({1-\delta})$ and the vertical arrow $\downarrow$ implies $n_2$ with value $-({1-\delta})$.\\
In this connection let us point out that usually in a two dimensional vortex  corresponding to spin $\frac {1}{2}$ quantum spin model the arrows $\rightarrow$ and  $\uparrow$ signify eigenstates of $S_x$ and $S_y$ with eigenvalues $+{\frac{1}{2}}$ respectively $^{5}$.  $\leftarrow$  and $\downarrow$ are eigenstates of these operators with eigenvalues $-{\frac{1}{2}}$ respectively. In our quantum formulation we have the spin coherent field components $n_1$ and  $n_2$ ( satisfying the constraint given by $Eqn. (4)$ at each lattice point ) forming the vortex (anti-vortex) with ${n_3}=\sin {\epsilon}$ at each lattice point. Our picture is that of a flattened "meron configuration" $^{6,7}$ mimicking a conventional  vortex or anti-vortex . Therefore $n_1$ or  $n_2$ cannot be exactly equal to $\pm 1$ . We choose them to be $\pm (1-\delta )$, where $\delta$ is a function of $\epsilon$ with $\delta \longrightarrow 0$ as  $\epsilon \longrightarrow 0$. Moreover if we want to identify the spin state $\vert \rightarrow \rangle$ (eigenstate of the operator $S_x$ with eigenvalue $+ {\frac{1}{2}}$) with the coherent state ${\cos{\frac{\theta}{2}}}{\vert  {\frac{1}{2}}\rangle} + ({e^{-\phi}}) {\sin{\frac{\theta}{2}}}{\vert  {-\frac{1}{2}}\rangle}$ at a vertex $[Fig.1]$ we have ${n_1}=1,{n_2}=0~and~{n_3}=0$ at that vertex. Similarly for $\vert \leftarrow \rangle$ we have ${n_1}=-1,{n_2}=0~and~{n_3}=0$ ; for $\vert \uparrow \rangle$ we have ${n_1}=0,{n_2}=1~and~{n_3}=0$ ; for  $\vert \downarrow \rangle$   ${n_1}=0,{n_2}=-1~and~{n_3}=0$ , in case we want to identify the respective spin states  with the coherent spin states. This does not agree with nonzero magnitude of    ${n_3}$ , given by ${n_3} = \sin{\epsilon}$ . Thus we choose ${n_1}$ or ${n_2}$ to be $\pm (1-\delta )$ to preserve the constraint given by $Eqn.(4)$.\\
Now it can be easily verified from $Eqn.(15)$ that ${{\bf \nabla}^2}{n_3} = 0$ at all lattice points. Then using $Eqns.$ $(11)$ and $(13)$ we can easily show that the contribution of  ${S_{WZ}^{tot}}$  to the plaquette in $Fig.1$ is
\begin{eqnarray}
{S_{{WZ}_1}}& = &{\int_0^{\beta}}dt  \{ [\lambda {n_2}{n_1}(3) + 2 {{{\sin}^2}{\epsilon}}~~ {n_2}(3)]{{\bf \nabla}^2}{n_2}(3)\nonumber\\
&   & + [-{\lambda}({{n_2}^2}(3) + {{\sin}^2}{\epsilon}) + 2 {{{\sin}^2}{\epsilon}}~~ {n_1}(3)]{{\bf \nabla}^2}{n_1}(3)\nonumber\\           
&   &+ [{n_1}(1){n_2}(2){n_1}(2) - {n_2}(1)[{{\sin}^2}{\epsilon}+{{n_1}^2}(2)]+  {{{\sin}^2}{\epsilon}}~~ {n_2}(2)]{{\bf \nabla}^2}{n_2}(2)\nonumber\\
&   &+ [-{n_1}(1)({{\sin}^2}{\epsilon}+{{n_2}^2}(2))+{n_2}(1){n_2}(2){n_1}(2)+{{{\sin}^2}{\epsilon}}~~ {n_1}(2)]{{\bf \nabla}^2}{n_1}(2)\nonumber\\
&   &+ [{n_1}(1){n_2}{n_1}(4) - {n_2}(1)({{\sin}^2}{\epsilon}+{{n_1}^2}(4)) + {{{\sin}^2}{\epsilon}}~~ {n_2}(4)]{{\bf \nabla}^2}{n_2}(4)\nonumber\\
&   &+ [-{n_1}(1)({{{\sin}^2}{\epsilon}}+{{n_2}^2}(4)) + {n_2}(1){n_1}{n_2}(4) + {{\sin}^2}{\epsilon}~~ {n_1}(4) ]    {{\bf \nabla}^2}{n_1}(4) \}   
\end{eqnarray}
Using $Eqns.(15)~~and~~(16)$ the derivatives on the right hand side of $Eq.(18)$ are given by:
\begin{eqnarray}
{{\bf \nabla}^2}{n_2}(3) & = & 4{n_2}(3)\nonumber\\
{{\bf \nabla}^2}{n_1}(3) & = & 4{n_1}(3) - 2\kappa \nonumber\\
{{\bf \nabla}^2}{n_2}(2) & = & 4{n_2}(2) - 2\mu \nonumber\\
{{\bf \nabla}^2}{n_2}(2) & = & 4{n_1}(2)\nonumber\\
{{\bf \nabla}^2}{n_2}(4) & = & 4{n_2}(4) - 2\mu \nonumber\\
{{\bf \nabla}^2}{n_1}(4) & = & 4{n_1}(4)
\end{eqnarray}
where $\mu$ and $\kappa$ are given by 
\begin{eqnarray}
\mu & = & {n_2}(1) + {n_2}(3)\nonumber\\
\kappa & = & {n_1}(2) + {n_1}(4)
\end{eqnarray}
Notice that while obtaining $Eq.(18)$ we have assumed the local PBC [$Eqn.(16)$]
so that the derivative ${{\bf \nabla}^2}{\bf n}$ at a point on the plaquette [$Fig.1$] is written in terms of the fields $\bf n$ at points on the same plaquette. In $Eq.(18)$ we have also made use of the fact that ${n_1}(1)+{n_1}(3) = (1-{\delta})-(1-{\delta})=0$ and  ${n_2}(2)+{n_2}(3) = (1-{\delta})-(1-{\delta})=0$ in [$Fig.1$]. Now we subtitute $Eq.(18)$ into $Eq.(17)$ and write $S_{{WZ}_1}$ as $A+B$. It is interesting to note that $A$ remains the same if we go from vortex to antivortex by changing  ${n_2}(2)$ and ${n_2}(4)$ in $Fig.1$ to $-{n_2}(2)$ and $-{n_2}(4)$ respectively whereas $B$ goes over to $-B$ . Thus  $S_{{WZ}_1}$ takes the form $A-B$ for antivortex. We have the following expression for ${S_{{WZ}_1}}$ for a 1-vortex .  
\begin{eqnarray}
{S_{{WZ}_1}} & = & A + B \nonumber\\
A & = &{\int_0^{\beta}}dt      \{ 8{{\sin}^2}{\epsilon}~{{\cos}^2}{\epsilon}  \nonumber\\
&   & + 2   {\mu}{n_2}(1)[2{{\sin}^2}{\epsilon}+{{n_1}^2}(2)+{{n_1}^2}(4)]+4{{\sin}^2}{\epsilon}[{{n_2}^2}(2)+{{n_2}^2}(4)] \}\nonumber\\
B & = & -4{\int_0^{\beta}}dt~~{\mu}~{n_1}(1){n_1}(2){n_2}(3)\cdot {n_2}(2)
\end{eqnarray}
where we have used the symmetry property that ${n_2}(2)  = - {n_2}(4)$ and further the constraint ${{\bf n}^2}=1$ at every site. It should be noted that the entire scheme is meaningful only if $B$ is nonzero. The case $B=0$ corresponds to the configurations in $Fig.~1$ with  ${n_2}(1) = {n_1}(2) =  {n_2}(3) = {n_1}(4) = 0$ . We consider only those configurations in $ Fig.~1$ in which none of the above equations is satisfied . The $WZ$ term  distinguishes between vortex or antivortex for such configurations only , corresponding to the above case . We will see more such $ B = 0 $ configurations corresponding to higher topological charges , for which $WZ$ term is not sensitive.   

:{\bf Analysis of 2-vortex }:\\
For a typical 2-vortex we refer to $Fig.2$ . We first calculate the contribution of the WZ-term on such a plaquette by simply adding the contributions of WZ from  each of the individual elementary plaquettes (subvortices) viz., $a$ , $b$ , $c$ and $d$, each carrying topological charge $+1$ , belonging to the above 2-vortex. This we denote by ${[a + b+ c + d]}_{FREE}$.     
\begin{figure}[!htbp]
\begin{center}
\includegraphics[keepaspectratio,width=10cm]{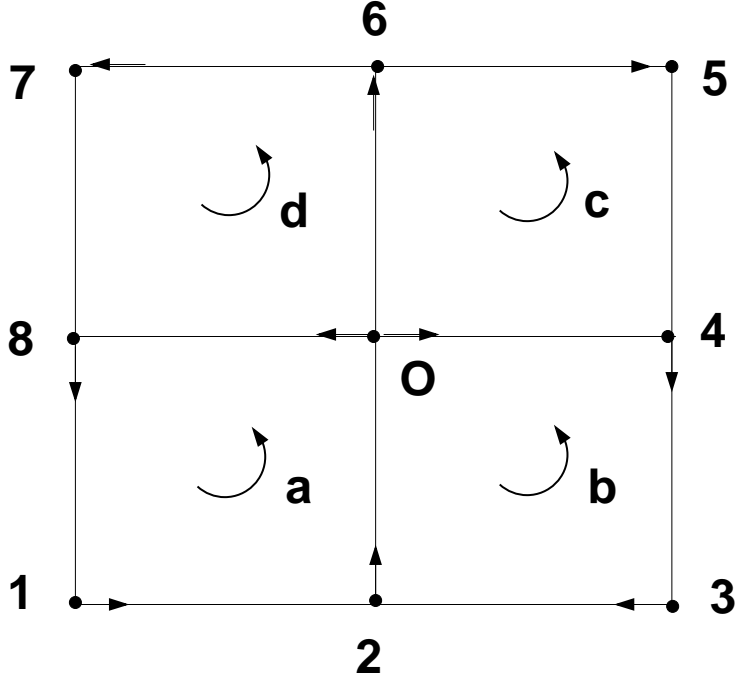}
\end{center}
\caption{2-vortex}
\end{figure}
We also calculate the contributions of the WZ-term from the above four subvortices as they are glued in a natural way to form the above mentioned 2-vortex [ see in $Fig.2$] .  We implement this by taking the contribution of the WZ-term with a factor of $\frac{1}{2}$ by taking into consideration the common bonds occuring between the pairs of adjacent subvortices. We denote this total contribution by ${[a+b+c+d]}_{GLUED}$. Let us now write down the formal expressions of  ${[a+b+c+d]}_{FREE}$ and  ${[a+b+c+d]}_{GLUED}$. 
\begin{eqnarray}
{{[a+b+c+d]}_{FREE}}& = & {\int_0^{\beta}}dt   \{ {\bf n}(1a)\cdot  ({\bf n}\wedge {\partial_t}{\bf n})(2a) + {\bf n}(2a)\cdot ({\bf n}\wedge {\partial_t}{\bf n})(3a)\nonumber\\
&   & + {\bf n}(4a)\cdot ({\bf n}\wedge {\partial_t}{\bf n})(3a) + {\bf n}(1a)\cdot ({\bf n}\wedge {\partial_t}{\bf n})(4a)\nonumber\\
&   & + {\bf n}(1b)\cdot ({\bf n}\wedge {\partial_t}{\bf n})(2b) + {\bf n}(2b)\cdot ({\bf n}\wedge {\partial_t}{\bf n})(3b)\nonumber\\
&   &+ {\bf n}(4b)\cdot ({\bf n}\wedge {\partial_t}{\bf n})(3b) + {\bf n}(1b)\cdot ({\bf n}\wedge {\partial_t}{\bf n})(4b)\nonumber\\  
&   &+ {\bf n}(1c)\cdot ({\bf n}\wedge {\partial_t}{\bf n})(2c) + {\bf n}(2c)\cdot ({\bf n}\wedge {\partial_t}{\bf n})(3c)\nonumber\\
&   &+ {\bf n}(4c)\cdot ({\bf n}\wedge {\partial_t}{\bf n})(3c) + {\bf n}(1c)\cdot ({\bf n}\wedge {\partial_t}{\bf n})(4c)\nonumber\\
&   &+ {\bf n}(1d)\cdot ({\bf n}\wedge {\partial_t}{\bf n})(2d) + {\bf n}(2d)\cdot ({\bf n}\wedge {\partial_t}{\bf n})(3d)\nonumber\\
&   &+ {\bf n}(4d)\cdot ({\bf n}\wedge {\partial_t}{\bf n})(3d) + {\bf n}(1d)\cdot ({\bf n}\wedge {\partial_t}{\bf n})(4d) \}
\end{eqnarray}
\begin{eqnarray}
{{[a+b+c+d]}_{GLUED}}& = &  {\int_0^{\beta}}dt   \{  {\bf n}(1a)\cdot ({\bf n}\wedge {\partial_t}{\bf n})(2a) +{\frac{1}{2}}{\bf n}(2a)\cdot ({\bf n}\wedge {\partial_t}{\bf n})(3a)\nonumber\\
&   & +{\frac{1}{2}}{\bf n}(4a)\cdot ({\bf n}\wedge {\partial_t}{\bf n})(3a) + {\bf n}(1a)\cdot ({\bf n}\wedge {\partial_t}{\bf n})(4a)\nonumber\\
&   & + {\bf n}(1b)\cdot ({\bf n}\wedge {\partial_t}{\bf n})(2b) + {\bf n}(2b)\cdot ({\bf n}\wedge {\partial_t}{\bf n})(3b)\nonumber\\
&   &+{\frac{1}{2}}{\bf n}(4b)\cdot ({\bf n}\wedge {\partial_t}{\bf n})(3b) +{\frac{1}{2}}{\bf n}(1b)\cdot ({\bf n}\wedge {\partial_t}{\bf n})(4b)\nonumber\\
&   &+{\frac{1}{2}}{\bf n}(1c)\cdot ({\bf n}\wedge {\partial_t}{\bf n})(2c) + {\bf n}(2c)\cdot ({\bf n}\wedge {\partial_t}{\bf n})(3c)\nonumber\\
&   &+ {\bf n}(4c)\cdot ({\bf n}\wedge {\partial_t}{\bf n})(3c) +{\frac{1}{2}}{\bf n}(1c)\cdot ({\bf n}\wedge {\partial_t}{\bf n})(4c)\nonumber\\
&   &+{\frac{1}{2}}{\bf n}(1d)\cdot ({\bf n}\wedge {\partial_t}{\bf n})(2d) +{\frac{1}{2}}{\bf n}(2d)\cdot ({\bf n}\wedge {\partial_t}{\bf n})(3d)\nonumber\\
&   &+ {\bf n}(4d)\cdot ({\bf n}\wedge {\partial_t}{\bf n})(3d) + {\bf n}(1d)\cdot ({\bf n}\wedge {\partial_t}{\bf n})(4d)  \}
\end{eqnarray}
With the help of $Eqs.(13)~ and~ (15)$, we now determine the following quantity $R$:
\begin{equation}
R= RELEVANT~~ PART \{ {{[a+b+c+d]}_{FREE}} - 2{{[a+b+c+d]}_{GLUED}} \}
\end{equation}
The RELEVANT PART contains the terms which are linear in ${n_2}(2)$,  ${n_2}(4)$, ${n_2}(6)$ and ${n_2}(8)$. They are called $RELEVANT$ because if we flip the directions of ${n_2}(2)$,  ${n_2}(4)$, ${n_2}(6)$ and ${n_2}(8)$ , we go from the state of vortex to its corresponding anti-vortex state [$see~~Fig.2$] .The quantity $R$ has been chosen in a way  so as to represent the degree of gluing amongst the constituent elementary plaquettes , while the composite object undergoes topological charge reversal. Thus the vanishing of $R$ provides a minimal condition for the lattice spin configuration on a plaquette to attain a topological character .\\
Now the quantity $R$ defined in $Eq.~(24)$ is given by
\begin{eqnarray}
R & = &{\int_0^{\beta}}dt  \{ {n_2}(2)[{n_1}(4) - {n_1}(6)]  [-2{n_1}(2){n_2}(1)+{n_1}(2){n_1}(3)\nonumber\\
&   & + {4{n_1}(1)}{{n_2}(3)}+ 2{{n_2}(0)}{{n_1}(1)}-{{n_2}(5)}{{n_1}(1)}] \}
\end{eqnarray}
where we have assumed for simplicity ${n_1}(2) = {n_1}(8)$. Thus the right hand side of the above equation vanishes if the following two conditions are satisfied
\begin{eqnarray}
{n_1}(2) & = & {n_1}(8)\nonumber\\
{n_1}(4) & = & {n_1}(6)
\end{eqnarray}
These conditions can easily be satisfied for a large class of spin configurations possessing topological charge 2.\\  
Henceforth we consider the equation
\begin{equation}
R = 0
\end{equation}
as the criterion for a spin configuration corresponding to a Q-vortex, to be a topological excitation , satisfying the  internal consistency .\\ 
For Q=3  [see~Fig.~3], for example ,we have 
\begin{equation}
R= RELEVANT~~ PART \{ {{[a+b+....+h]}_{FREE}} - 2{{[a+b+....+h]}_{GLUED}} \}
\end{equation}
with $Eqn.~(27)$ being the criterion for the 3-vortex to be a topological excitation .  

\vspace{0.2cm}
      
  Let us mention that in the present case for Q=2, there are configurations corresponding to the vanishing of the relevant term in the glued vortex i.e, $B=0$ , which fulfil the condition $ R = 0 $ as well in $Eqn.~(27)$. Our scheme is however applicable to only $ R=0 $ configurations having finite $B$ . The fact that the $WZ$ term is sensitive only to a particular class of configurations , has its origin in the presence of a non-topological part along with the topological one [$see~ Ref.~3$] , in it. \\

:{\bf Analysis of 3-vortex }:\\
\begin{figure}[!htbp]
\begin{center}
\includegraphics[keepaspectratio,width=10cm]{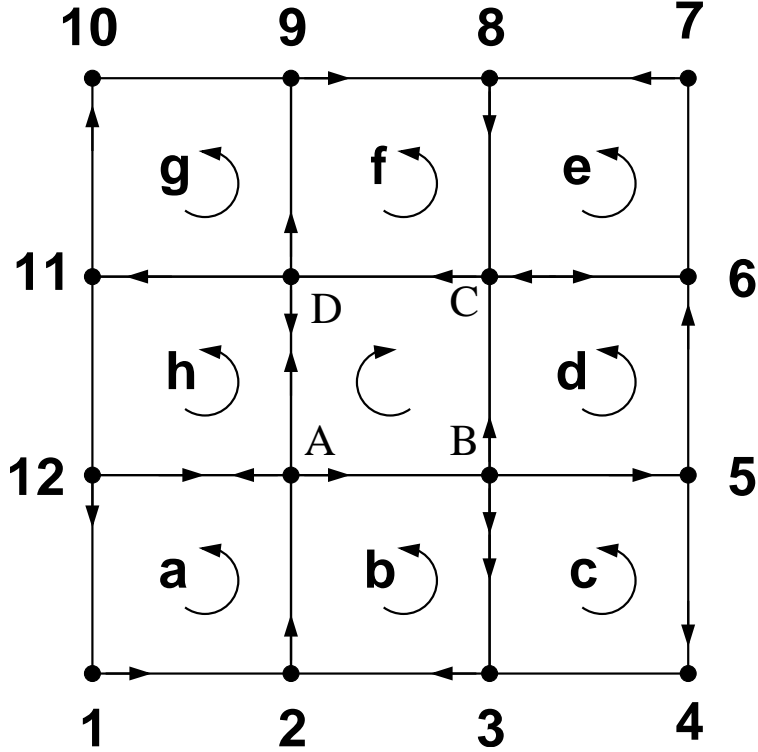}
\caption{3-vortex}
\end{center}
\end{figure}
We directly refer to the $Fig.~3$. We obtain the cofugurations corresponding to the following minimal conditions satisfying $Eqn.~(27)$

\begin{eqnarray}
{n_1}(2) & = & {n_1}(12)\nonumber\\
{n_1}(6) & = & {n_1}(8)\nonumber\\
{n_1}(4) & = & {n_1}(10)\nonumber\\
{n_1}(B) & = & {n_1}(D)\nonumber\\
{n_2}(1) & = & {n_2}(3)\nonumber\\
{n_2}(5) & = & {n_2}(9)\nonumber\\
{n_2}(3) & = & {n_2}(11)\nonumber\\
{n_2}(C) & = & {n_2}(D)
\end{eqnarray}

Here also we do not consider the configurations corresponding to $B=0$ in the glued vortex. Studying the configuration in  $Fig.~3$ we discover the following identity for a topological Q-vortex ,with odd Q ( $Q \ne 1$ ), which describes the  topological charge  ditribution inside the vortex  consistently.
\begin{equation} 
Q = {\frac{1}{4}} [{Q^2} - {{(Q - 2)}^2}] + 1
\end{equation}
The quantity $(Q-2)$ is the absolute value of the effective charge ( negative ) of the core which is an anti-vortex in this case. Notice that the above equation holds good even within the core, i.e., when $Q$ is replaced by $(Q-2)$, and thereby depicts a ${\bf self~similar~pattern}$.

:{\bf Analysis of 4-vortex }:\\
\begin{figure}[!htbp]
\begin{center}
\includegraphics[keepaspectratio,width=10cm]{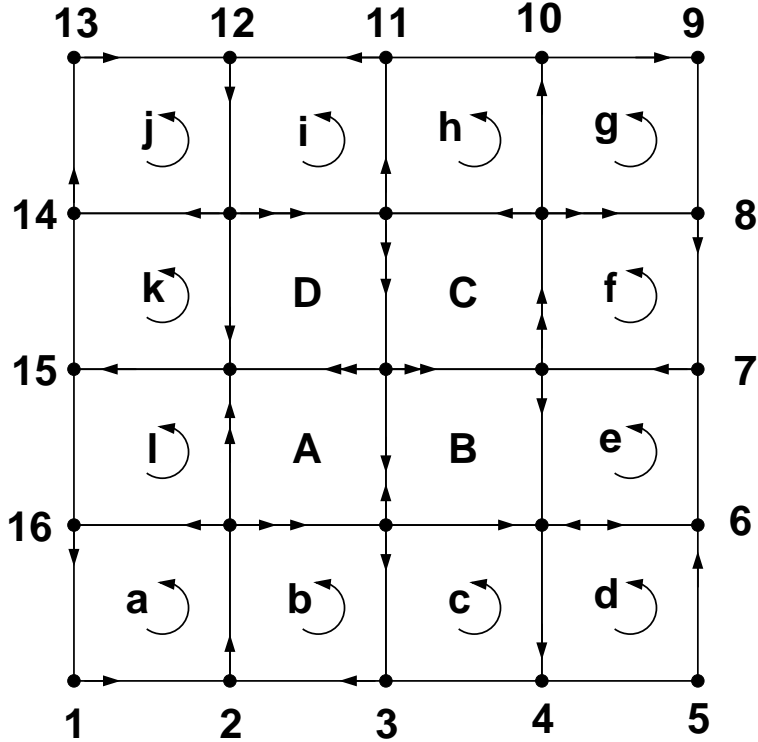}
\end{center}
\caption{4-vortex}
\end{figure}
In the case of 4-vortex $Fig.~4$ we have :
\begin{equation}
R = RELEVANT~~PART \{ {{[a+b+.....+k+l]}_{FREE}} - 2{{[a+b+.....+k+l]}_{GLUED}} \}
\end{equation}
and minimal solutions of $Eqn.~(27)$ as in the following :
\begin{equation}
{n_1}(even~points)  = {n_2}(odd~points ) = 0
\end{equation}  
 where even and odd points imply the vertices on the boundary of the 4-vortex with even and odd integer labels. This solution $[Eqn.(32)]$ also corresponds to vanishing  of the term $ B = 0 $ for the glued vortex. Moreover, the self similarity pattern, as discussed in $Eqn.~(30)$ , breaks down in this case. \\
The above two features continue to hold for vortices ( anti-vortices ) of higher even-valued topological charges .

{\bf Conclusion and Discussion}\\
i) Our calculations and analysis with $2D$ spin $\frac {1}{2}$ ferromagnetic quantum $XY$ model clearly bring out distingushing features between the even and odd charge sectors. The charge distribution equation $[Eqn.~(30)]$ involving the full spin configuration, is obeyed by the odd charge excitation in totality. The even charge sector however is not governed by this equation.
  
ii) Our work has established the role of WZ term as a topological charge measuring quantity , obtained from microscopic theory. Furthermore, this term is able to test various internal consistencies and conservation conditions involving the topological charge distribution for a large class of composite excitations . Thus our approach is more powerful than that based on the heuristic operators  suggested for determining the  charges of topological excitations $^{5,8}$. 

iii) The restriction on the class of excitations identifiable by $WZ$ term  on a lattice , arises from the presence of a non-topological contribution , besides the usual topological one$^3$.

iv) Our future plan includes the generalization of our approach to the case of finite $\lambda_z$ to achieve a physical realization of meronic type of excitations in the spin models . Furthermore making use of these results , we aim to evaluate the static and dynamic spin correlations for two dimensional spin $\frac {1}{2}$ anisotropic quantum Heisenberg ferromagnet at any temperature . This can also throw some light on the possible phenomenological scenario of quantum Kosterlitz - Thouless transition $^{3,9}$.\\
 To conclude, this study of topological spin excitations on the 2D-lattice will undoubtedly play an important role in the understanding of thermodynamics of low dimensional ferromagnets.\\ 
{\bf Acknowledgement} : One of the authors (SKP) would like to thank M.G.Mustafa, Rajarshi Roy and Purnendu Chakraborti for their valuable help in the preparation of the figures.

\vspace{0.1cm}
 
 We dedicate this piece of work to the memory of Late Professor Chanchal Kumar Majumdar.

{\bf Reference}\\
$^1$E. Fradkin and M. Stone , Phys. Rev. B {\bf 38}, 7215 (1988).\\  
$^2$E. Fradkin, $Field$ $Theories$ $of$ $Condensed$ $Matter$ $Systems$ ( Addision-Wesley, CA, 1991).\\
$^3$Ranjan Chaudhury and Samir K. Paul , Phys. Rev. B {\bf 60} 6234 (1999).\\
$^4$Ranjan Chaudhury and Samir K. Paul , Mod. Phys. Letts. B {\bf 16} 251 (2002).\\
$^5$E. Loh, Jr., D. J. Scalapino and P. M. Grant , Phys. Rev. B {\bf 31} 4712(1985) ;  R. H. Swendsen, Phys. Rev. Lett. {\bf 49}, 1302 (1982) ;  D. D. Betts, F. C. Salevsky and J. Rogiers, J. Phys. A {\bf 14} (1981).\\ 
$^6$Mona Berciu and Sajeev John , Phys. Rev. B {\bf 61} 16454 (2000).\\
$^7$T. Morinari , J. Magn. Mag. Mat. {\bf 302} , 382 (2006).\\ 
$^8$ A. S. M\'ol , A. R. Pereira , H. Chamati and S. Romano , Eur. Phys. J. B {\bf 50} , 541 (2006).\\
$^9$A.Cuccoli , T. Valerio , V. Paola and V. Ruggero , Phys. Rev. B, {\bf 51} 12840 (1995). 
\end{document}